\documentclass[12pt]{article}  
\usepackage{amsmath}
\usepackage{url}
\usepackage{graphicx}
\usepackage{wrapfig}
\usepackage{lscape}
\usepackage{float}
\usepackage{rotating}
\usepackage{epstopdf}
\usepackage{setspace} 
\usepackage[font=footnotesize]{caption}
\usepackage{scicite}

\newenvironment{sciabstract}{%
\begin{quote} \bf}
{\end{quote}}

\newcounter{lastnote}

\title{Stock fluctuations are correlated and amplified across networks of interlocking directorates}
\author{Serguei Saavedra$^1$\footnote{To whom correspondence should be addressed. E-mail: serguei.saavedra@ebd.csic.es. Tel. +34 954 466 700 (ext. 1330), Fax. +34 954 621 125.}, Luis J. Gilarranz$^1$, Rudolf P. Rohr$^{1,2}$, \\Michael Schnabel$^{3,4}$, Brian Uzzi$^{3,4}$ and Jordi Bascompte$^1$
\\        
\\$^1$Integrative Ecology Group \\ Estaci\'on Biol\'ogica de Do\~nana (EBD-CSIC) \\ Calle Am\'erico Vespucio s/n \\ E-41092 Sevilla, Spain\\
\\$^2$Unit of Ecology and Evolution \\ Department of Biology, University of Fribourg \\ Chemin du Mus\'{e}e 10 \\ CH-1700 Fribourg, Switzerland\\
\\$^3$Northwestern Institute on Complex Systems, \\$^4$Kellogg School of Management \\ Northwestern University \\ Evanston, Illinois, 60208, USA.
}

\date{}

\begin{document}
\maketitle

\newpage

%\begin{linenumbers}

\begin{sciabstract}
Traded corporations are required by law to have a majority of outside directors on their board. This requirement allows the existence of directors who sit on the board of two or more corporations at the same time, generating what is commonly known as interlocking directorates. While research has shown that networks of interlocking directorates facilitate the transmission of information between corporations, little is known about the extent to which such interlocking networks can explain the fluctuations of stock price returns. Yet, this is a special concern since the risk of amplifying stock fluctuations is latent. To answer this question, here we analyze the board composition, traders' perception, and stock performance of more than 1500 US traded corporations from 2007-2011. First, we find that the fewer degrees of separation between two corporations in the interlocking network, the stronger the temporal correlation between their stock price returns. Second, we find that the centrality of traded corporations in the interlocking network correlates with the frequency at which financial traders talk about such corporations, and this frequency is in turn proportional to the corresponding traded volume. Third, we show that the centrality of corporations was negatively associated with their stock performance in 2008, the year of the big financial crash. These results suggest that the strategic decisions made by interlocking directorates are strongly followed by stock analysts and have the potential to correlate and amplify the movement of stock prices during financial crashes. These results may have relevant implications for scholars, investors, and regulators.
\end{sciabstract}

%\clearpage


\baselineskip=8.5mm

\section{Introduction}

According to corporate governance standards in the US \cite{NYSE}, traded corporations are required to have a minimum of three directors in their board and a majority of outside directors \cite{Mizruchi,Burt}. Typically, these outside directors either have their primary affiliation with a different corporation, are self-employed, or retired \cite{Mizruchi}. This allows a director to sit on the board of two or more corporations at the same time---including or not their own affiliation board, generating what is commonly known as interlocking directorates \cite{Dooley,Mizruchi,Burt,Carpenter}. In turn, interlocking networks emerge as the result of many interconnected interlocking directorates \cite{Mizruchi,Burt}. Indeed, interlocking networks have been the focus of numerous research analyzing their role in the performance of corporations, organizational failure, economic downturns, hegemony, CEO pay, the sale price of a corporation, stock synchronicity, and prices for a corporation's services, among others \cite{Burt,Mizruchi,Pennings,Palmer,Fligstein92,Fligstein,Khanna}. Yet, the strongest consensus so far is that interlocking networks mainly favor the transmission of information among corporations \cite{Burt,Borgatti,Carpenter}. For instance, this information advantage has been observed when central corporations of interlocking networks can adopt new market strategies quicker via information they gather from their interlocking directorates \cite{Connelly,Davis}.

A major question that remains to be answered is whether interlocking networks can affect market processes \cite{Burt}. Board members are typically major stockholders and report financial strategies to their investors \cite{Burt,Mizruchi}. Indeed, previous work has suggested that interlocking directorates may play an important role in the information gathered by stock analysts, especially when there is substantial media coverage \cite{Haunschild}. This is a special concern of regulators since the risk of amplifying stock fluctuations is latent \cite{Burt,Mizruchi}. In fact, this question has motivated important actions in the US Congress dating back to the early 1900's \cite{Mizruchi}. Section 8 of the Clayton Act explicitly prohibited interlocks if the linked corporations would violate antitrust laws if combined into a single corporation \cite{Dooley,Mizruchi}. Over the years, these actions have promoted a decrease in the number of interlocking directorates, especially within the financial sector \cite{Burt}. However, even with these limitations, it has been shown that financial shocks can be transmitted across the interlocking network regardless of the degrees of separation between corporations \cite{Burt,Mizruchi,Vedres,Kogut,Battiston}.

To shed new light on the above question, we study the extent to which interlocking networks can explain stock fluctuations. We analyze data on the board composition, traders' perception, and stock performance of more than 1500 US traded corporations (see Appendix for further details). To capture the behavior of interlocking networks across a variety of financial periods, we focus our analysis on the period covered by the big 2008 financial crash and its pre and post financial periods. We use the data extracted from {\it RiskMetrics} \cite{RiskMetrics}, which is a yearly compilation from 2007-2011 of the board composition of over 1500 large US traded corporations. Daily closing stock prices for these corporations are extracted from {\it WRDS} database \cite{WRDS}, and stocks are categorized in a market sector according to {\it Yahoo Finance} criteria \cite{Yahoo}. We analyze traders' perception on these traded corporations using trading data and electronic communication among traders from one US trading firm for the period 2007-2009 \cite{Saavedra,Saavedra2}. Here, we explore the extent to which interlocking networks can explain stock correlations, the plausible mechanisms linking interlocking networks and stock markets, and the potential effect of such associations.

\section{Results}

\subsection{Network characterization}

We constructed interlocking networks for each year from 2007-2011 formed only by the traded corporations observed in a particular year. A link between two corporations is established if they share at least one board member in the same year. Consistent with previous studies and regulations \cite{Mizruchi,Burt}, we find that the observed boards in these interlocking networks have a median size of $9$ directors, of which the majority are outside directors ($>50\%$). The $99\%$ of the observed interlocking directorates (i.e., links between two corporations) share only one board member. In total, there is approximately $20\%$ of corporations who do not share any single director with another corporation, whereas the other $80\%$ are on average $4.6$ interlocking directorates apart from each other. Interestingly, $85\%$ of all interlocking directorates are formed between corporations that belong to different market sectors, and we find no significant connectivity differences across market sectors. These results are also in agreement with the characterization of interlocking networks found in previous studies \cite{Kogut,Battiston}.

Whereas the characterization of the interlocking network remains fairly constant across the observation period, the financial market exhibited important fluctuations from year to year. In 2007, $46\%$ of the 1422 observed traded corporations increased their stock price by the end of the year. In contrast, in 2008, the year of the big financial crash, only $12\%$ of the 1453 observed traded corporations increased their stock price. This was followed by a recovery period in 2009 and 2010 were $81\%$ of 1357 and 1413 observed corporations increased their stock price, respectively. Finally, in 2011, the market had a relatively bad period with $41\%$ of 1447 observed corporations increasing their stock price. Therefore, these data bring us the opportunity to investigate the association between interlocking networks and stock fluctuations across different financial periods.

\subsection{Interlocking networks and stock correlations}

To study whether stock correlations can follow a characteristic pattern among interlocking directorates, we measure the association between network proximity---degrees of separation---and market similarity---temporal correlation of stock prices---among traded corporations (Fig. 1). It was previously shown that the synchronicity between two stocks can increase if these two corporations share a director \cite{Khanna}. Here, we expand on this question to ask whether traded corporations that are closer in the interlocking network also have a stronger correlation between their stocks. For each year, we define a matrix of network proximity $\boldsymbol{D}$ of size $N\times N$, where $N$ is the number of observed traded corporations in a year, $D_{ij}=1/(d_{ij})$, and $d_{ij}$ is the degree of separation (number of links) between corporation $i$ and $j$ in the interlocking network. Note that the degree of separation of a corporation to itself is $d_{ii}=0$, and unconnected corporations have an infinite degree of separation ($d_{ij}=\infty$). The greater $D_{ij}$, the higher the proximity between the two corporations in the interlocking network. Note that our matrix of network proximity takes into account all pairs of traded corporations.

Additionally, we define a matrix of market similarity $\boldsymbol{S}=\frac{1}{M-1}RR^{T}$ of size $N \times N$, where $M$ is the total number of trading days in a year, and $R_{ik}=(z_{ik} - \langle z_{ik} \rangle )/\sigma_{z_{ik}}$ is the standard deviation normalized daily log returns of stock $i$ in day $k$ such that $z_{ik}=\text{log}(p_i(k)/p_i(k-1))$ and $p_i(k)$ is the closing stock price of corporation $i$ in day $k$. Therefore, the matrix of market similarity $\boldsymbol{S}$ corresponds to the correlation matrix $\Sigma _{R}$ of $R$ \cite{Fenn}. The higher $S_{ij}$, the higher the similarity or temporal correlation of the stock price movements of corporation $i$ and $j$ in the year. 

We calculate partial Mantel correlations \cite{Mantel} (both Spearman and Pearson correlations) between the matrices of network proximity and market similarity while controlling for other proximity matrices given by market sector, geographic distance, board size, fraction of directors with financial expertise, and average stock price in the year (see Appendix for further details). Following previous studies \cite{Banz,Schwert,Cheung,DavisLJ,Berk}, we use the average stock price in the year as surrogate of firm size given its relevance and undisputed effect on the analysis of stock fluctuations. We also divide our analysis into the biggest market sectors (Basic Materials, Consumer Goods, Financial, Health Care, Industrial Goods, Services, and Technology), where sector matrices are simply sub-matrices of the full proximity matrices composed of corporations from a single sector.

Importantly, we find positive correlations between network proximity and market similarity in both individual sectors and in the market as a whole (i.e., taking all sectors together). Figure 2 shows that the majority (23 out of 35, $P=0.024$, binomial test) of partial Mantel correlations yield positive correlations with $95\%$ bootstrap confidence intervals (solid circles). These correlations hold to additional non-parametric statistical tests (see Appendix). While the correlations found have relatively small values (the largest is $0.4$), let us not forget that even small changes in the price of stocks can generate big losses or gains in the market \cite{Saavedra2,Moro}. Moreover, these correlation values are larger than the values generated if we only focus on distances $d_{ij}=1$ (see Appendix), following previous work \cite{Khanna}. Interestingly, we can also observe that not all sector behave in the same way. For instance, the interlocking networks of the healthcare and industrial sectors can explain more clearly the stock correlations among their constituent traded corporations than the consumer or financial sector. We can also see that some sectors, such as the financial or technological sector, change from negligible (positive) correlations at the beginning of the observational period to positive (negligible) correlations by the end of the period. We leave future explanations of these changes to the reader.

To further support the validity of the correlations between network proximity and market similarity, we test whether corporations that increase or decrease their network proximity from one year to the next one also increase or decrease their market similarity accordingly. In specific, for each year from 2008-2011, we calculate a new matrix of network proximity given by the difference between the network proximity between two corporations in a given year $D_{ij}(t)$ and their network proximity in the previous year $D_{ij}(t-1)$ (only taking into account those corporations that are present in both years). Similarly, we generate a new matrix of market similarity for each year from 2008-2011, in which the new elements of each of these matrices are given by the difference between the market similarity between two corporations in a given year $S_{ij}(t)$ and their market similarity in the previous year $S_{ij}(t-1)$. Finally, for each year, we measure the correlation between the newly generated matrices of network proximity and market similarity. 

Figure 2 (green/Y region) shows that the new correlations between proximity changes are, in fact, equivalent to the positive correlations in each year. Overall, these findings reveal that changes in the degree of separation between two corporations in the interlocking network are correlated with changes in their stock correlations. If the degree of separation decreases, the stock correlation increases, and vice versa. 

\subsection{Interlocking networks and stock markets}

To unveil potential mechanisms explaining the previous association between stock correlations and interlocking networks, we study the behavior of one the main financial actors towards interlocking directorates. In specific, we analyze the perception and information gathered by stock analysts on traded corporations and whether this is associated with their trading activity. As proxy for the information collected by stock analysts, we use trading data and the electronic communication among a group of financial traders from a US trading company from 2007-2009 (see Appendix). Previous work has revealed that traders are constantly tracking business press coverage of traded corporations and exchanging this information among their peers \cite{Saavedra}. An illustrative example of this information concerning corporate directorates is the following message between two traders on July $7^{th}$ 2008: ``Microsoft willing to enter talks if Yahoo elects new board." In fact, it has been shown that traders' communications can signal their understanding of market volatility \cite{Saavedra,Saavedra2}. Therefore, this suggests that the more strategic decisions made by interlocking directorates, the higher the chances that there is a media coverage of relevance to traders, and the higher the potential that this correlates with their trading activity.

To test the above hypothesis, for each year from 2007-2009, we investigate the association of the frequency at which traders talk about traded corporations with the centrality in the interlocking network and the traded volume of such corporations. The frequency is calculated by the total number of times a ticker (e.g. GOOG as for Google) is mentioned  over the entire year by this group of traders. We measure a corporation's centrality by $\langle D_i \rangle = \frac{1}{N-1}\sum _j D_{ij}$, where $D_{ij}$ is the network proximity between two corporations, as mentioned above. We find that similar results are obtained if we replace $\langle D_i \rangle$ by other centrality measures, such as the total number of interlocking directors or the community participation coefficient \cite{Guimera}. The traded volume is calculated by the total amount of US dollars traded by this group of traders over the entire year. 

Figure 3 shows indeed a positive association (partial Spearman rank correlation) of ticker mentions with both the centrality and traded volume of corporations, controlling for the average stock price, board size, and fraction of financial experts in the board. Interestingly, we find negligible associations (partial Spearman rank correlation) between centrality and traded volume, controlling for ticker mentions, the average stock price, board size, and fraction of financial experts in the board. This supports the hypothesis of a path going from interlocking directorates to traders reacting to information and then to trading activity. Overall, these findings show an enhanced attention of stock analysts to central corporations in interlocking networks.

\subsection{Exposure to market fluctuations in interlocking networks}

Finally, to test the extent to which interlocking networks can amplify the exposure of traded corporations to market fluctuations, we compare the centrality of corporations in the interlocking network with their stock performance over the year. The centrality of a corporations is again measured by  $\langle D_i \rangle $. We measure a corporation's short-term and long-term stock performance by its beta ($\beta_i$) and its yearly stock price return ($r_i$), respectively. 

The commonly known beta of a stock is given by $\beta_i=\text{Cov}(z_i,z_b)/\text{Var}(z_b)$, where $z_i$ and $z_b$ are the daily log returns of stocks $i$ and the benchmark return, respectively (see Appendix). The higher (lower) the beta, the more the stock moves in the same (opposite) direction and farther apart from the benchmark return. This means that during good or bad market periods, the short-term stock performance of a traded corporation increases the higher or the lower the beta of the stock, respectively. The yearly stock price return is given by $r_i=\text{log}(p_i(t_f)/p_i(t_o))$, where $p_i(t_o)$ and $p_i(t_f)$ are the daily closing stock prices of corporation $i$ at the beginning and at the end of the calendar year, respectively. The higher the yearly return, the better the long-term stock performance of a traded corporation. For each year from 2007-2011, we measure the effect of centrality on the short-term and long-term stock performance using a multivariate linear regression model controlling for the corporations' individual characteristics (see Appendix).

In general, Figure 4 shows that the centrality of traded corporations has negligible effects on their long-term and short-term stock performance (but see the Financial sector). This is in line with our previous findings showing that interlocking networks favor stock correlations and, in turn, this may explain why interlocking networks have no effect on pushing traded corporations away from the general market trend. However, Figure 4 also reveals that during 2008, the year of the big financial crash, there is a negative effect between centrality and long-term stock performance. These results suggest that during financial crashes, interlocking networks can amplify the exposure of traded corporations to market fluctuations.

\section{Discussion}

US governance standards favor the participation of outside directors in corporate boards, which in turn allows the creation of interlocking directorates. While direct competitors are prohibited to create interlocking directorates, these corporations have on average only four degrees of separation. The financial sector is a good example of this market environment. There is almost no interlocking directorate between financial corporations, but they are indirectly linked by their interlocking directors sitting on the board of a third corporation belonging to a different market sector. Indeed, as we mentioned before, $85\%$ of interlocking directorates are formed between corporations belonging to different market sectors. While interlocking networks facilitate the transmission of information, it has been unclear whether they can also explain stock fluctuations in the financial market.

Our findings have shown that interlocking directorates seem to be one of the factors favoring the existence of stock correlations. We acknowledge that other confounding factors can explain stock correlations. Interestingly, a potential mechanism explaining the link between interlocking networks and stock markets appears to be the enhanced attention of stock analysts to central corporations. It remains to be seen whether other measures of centrality can provide better insights about these associations. Importantly, because interlocking networks may amplify market fluctuations during financial crashes, future work should also explore the impact of the different dynamics found across market sectors. Importantly, it does not escape our notice that these results provide valuable non-opinionated insight for scholars, regulators, investors, and board members themselves.
\\ 

\textbf{ACKNOWLEDGMENTS} Funding was provided by the European Research Council through an Advanced Grant (JB), the Spanish Ministry of Education through a FPU PhD Fellowship (LJG), FP7-REGPOT-2010-1 program under project 264125 EcoGenes (RPR), and the Kellogg School of Management (MS).
\\ \\
\textbf{Author contributions} All authors contributed extensively to the work presented in this paper.
\\
\textbf{Competing financial interests} The authors declare no competing financial interests.

%\end{linenumbers}

%\singlespacing

%References and Notes
\clearpage
\medskip

\renewcommand{\baselinestretch}{1.5}
{\small
\bibliographystyle{pnas}
\bibliography{bibliography}

}

\clearpage

%Figure Captions
\begin{figure}[ht]
\centerline{\includegraphics*[width= 0.5 \linewidth]{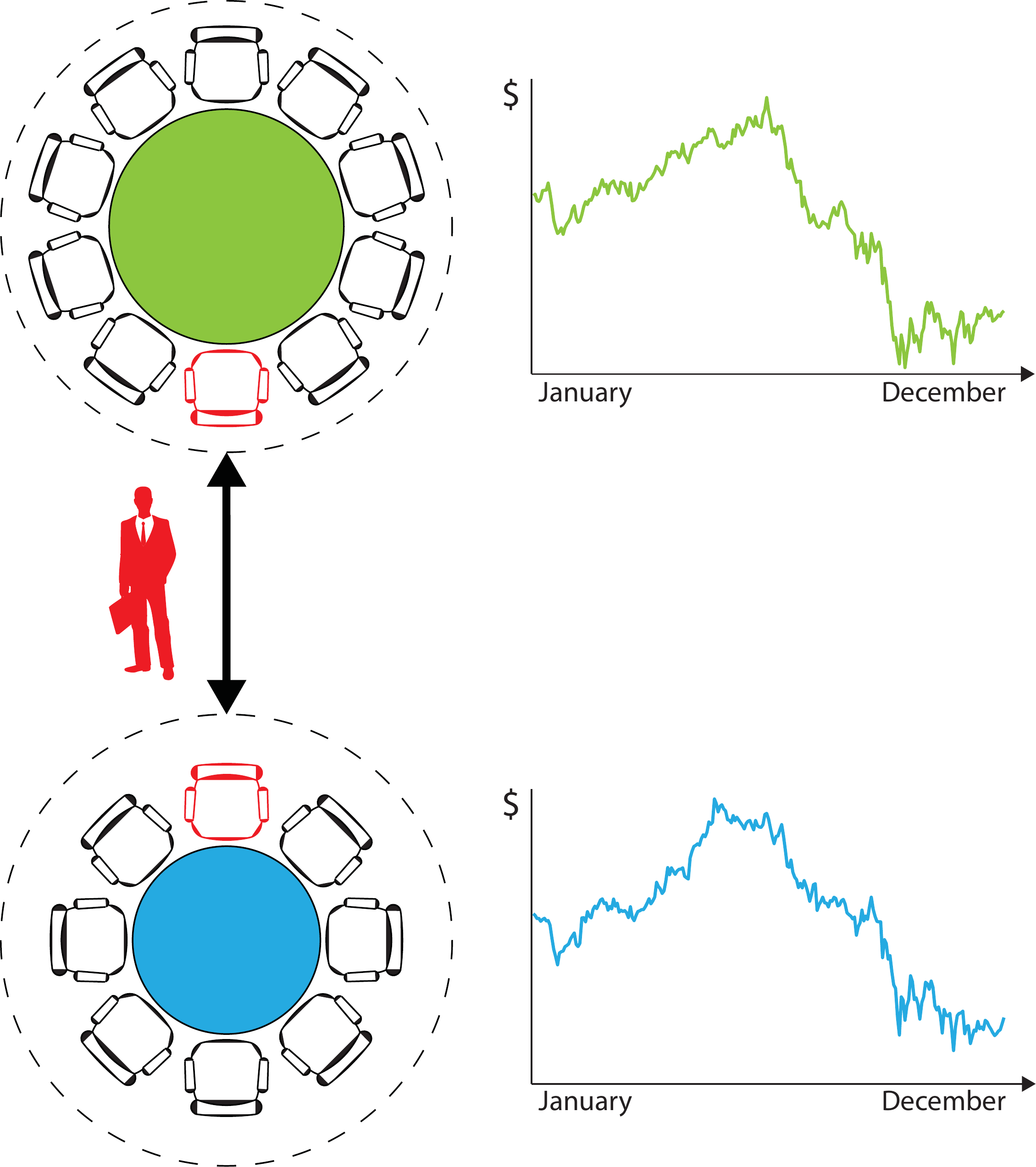}}
\scriptsize
\caption{\small Illustrative example of interlocking directorates. The green and blue traded corporations are linked by their inside/outside director that sits on their board. The question is whether traded corporations that are closer in the interlocking network also have a stronger correlation between their stocks.
}
\label{fig1}
\end{figure}

\clearpage

\begin{figure}[H]
\centerline{\includegraphics[width=1.5 \linewidth]{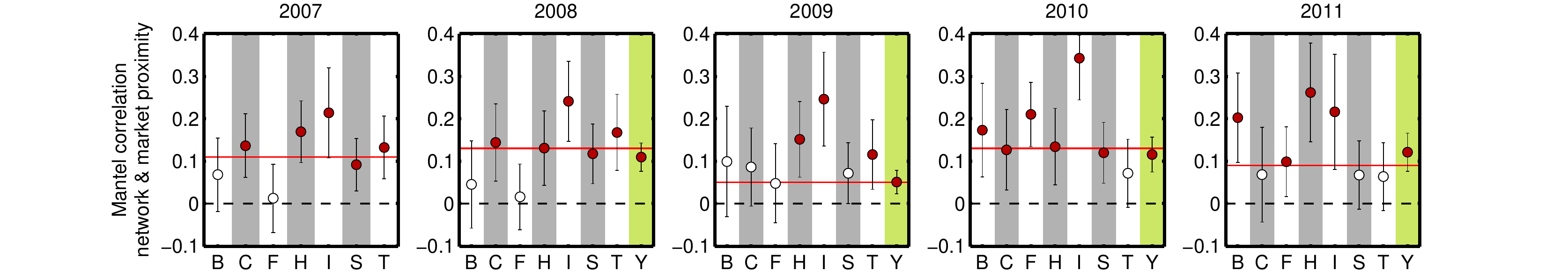}}
\caption{\small Interlocking networks and stock correlations. Spearman partial Mantel correlations (Pearson correlations yield similar results) between the matrices of network proximity and market similarity of traded corporations in each year. The red/solid line corresponds to the calculated correlation for all traded corporations regardless of market sector. Circles correspond to correlations in a particular market sector. Solid symbols correspond to correlations that do not cross zero using bootstrap $95\%$ confidence intervals (error bars). We focus on the seven major sectors: (B) basic materials, (C) consumer goods, (F) financial, (H) healthcare, (I) industrial, (S) services, and (T) technology. The Y column in 2008-2011 corresponds to the correlation between changes in network proximity and changes in market similarity for all pairs of traded corporations in reference to their values in the previous year. 
}
\label{fig2}
\end{figure}

\clearpage

\begin{figure}[H]
\centerline{\includegraphics[width=0.65 \linewidth]{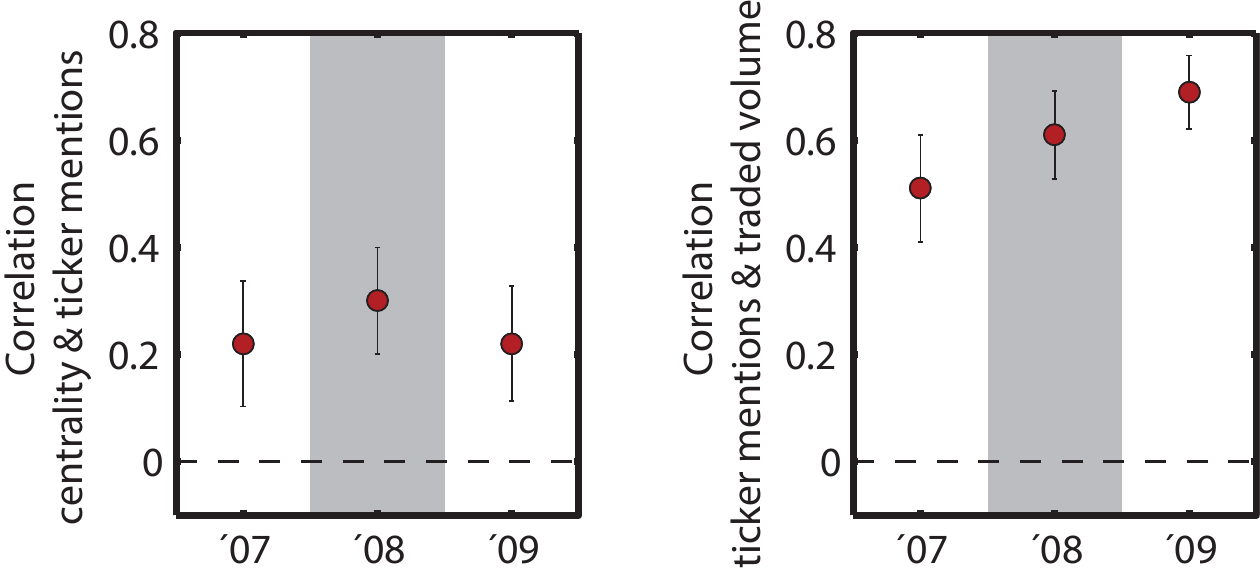}}
\caption{\small Interlocking networks and stock markets. For each year from 2007-2009, the figure shows Spearman rank correlations (Pearson correlations yield similar results) between tickers' mentions and the centrality of such corporations in the interlocking network. Similarly, the figure shows Spearman rank correlations (Pearson correlations yield similar results) between tickers' mentions and the traded volume of such corporations by the same group of traders over the entire year. 
}
\label{fig3}
\end{figure}

\clearpage

\begin{figure}[H]
\centerline{\includegraphics[width=1.4 \linewidth]{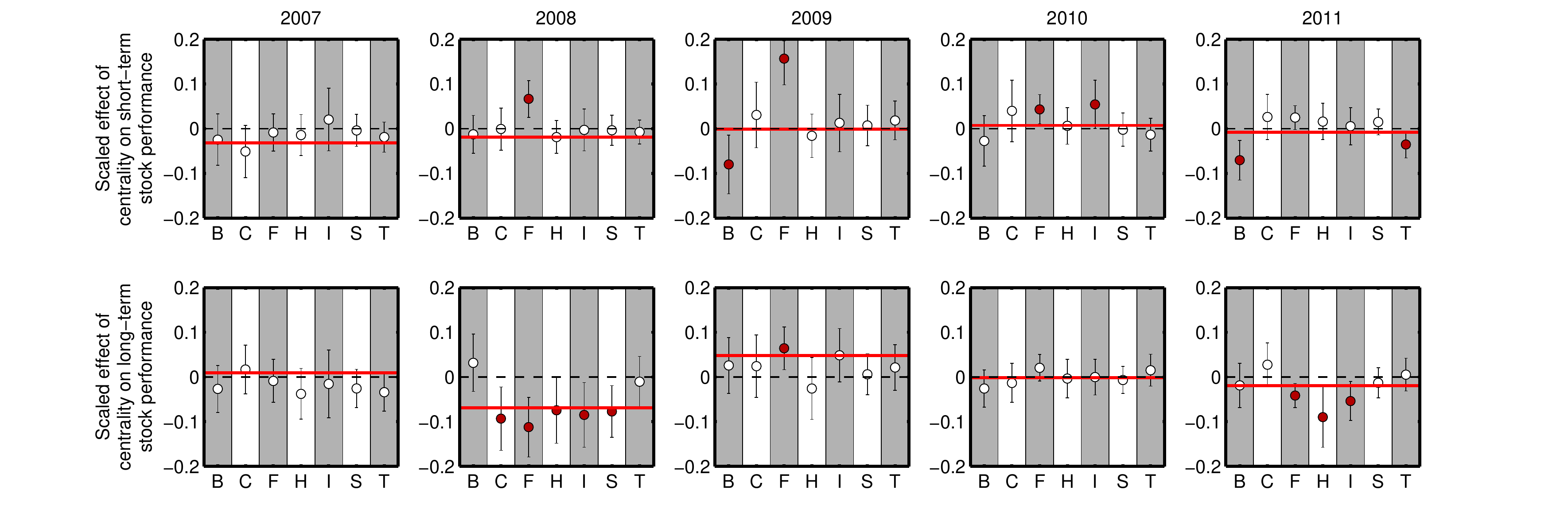}}
\caption{\small Exposure of traded corporations to market fluctuations. The first row corresponds to the scaled effect of centrality of traded corporations on their short-term stock performance---beta of stocks. The second row corresponds to the scaled effect of centrality of traded corporations on their long-term stock performance---yearly log returns of stocks. The red/solid line corresponds to the effect when taking into account all traded corporations regardless of market sector. Circles correspond to effects in a particular market sector. Solid symbols correspond to effects that do not cross zero using bootstrap $95\%$ confidence intervals (error bars). We focus on the seven major sectors: (B) basic materials, (C) consumer goods, (F) financial, (H) healthcare, (I) industrial, (S) services, and (T) technology.
}
\label{fig4}
\end{figure}

% SI labels have an S attached
\renewcommand{\thefigure}{S\arabic{figure}}
\renewcommand{\thetable}{S\arabic{table}}
\renewcommand{\thesection}{S\arabic{section}}
\renewcommand{\theequation}{S\arabic{equation}}
\renewcommand{\figurename}{Fig.}
%\renewcommand{\tablename}{Table}

% restart the figure counter
\setcounter{figure}{0}

\clearpage

%Supplementary Information

%\begin{linenumbers}

\section*{Appendix}

\textbf{Board composition data.} We use the data extracted from {\it RiskMetrics} \cite{RiskMetrics}, which is a yearly compilation from 2007-2011 of the board composition of over 1500 large US traded corporations. We use only those traded corporation on which we could find stock price information in our stock price data. Board members were manually disambiguated using gender, age, and affiliation data provided by {\it RiskMetrics}. These data also have information of whether a board member is consider a financial expert by the traded corporation. In some cases, the same board member is classified as a financial member in one corporations but a non-financial expert in a second corporation. We decided to classify board members as financial experts if they are considered to be experts at least in one corporation. The board composition and their financial expertise for each year is provided in an additional file.
\\
\textbf{Stock price data.} Daily closing stock prices from 2007-2011 for these corporations are extracted from {\it WRDS} database \cite{WRDS}, and stocks are categorized in a market sector according to {\it Yahoo Finance} criteria \cite{Yahoo}. The stock data and market classification for each year is provided in an additional file.
\\
\textbf{Traders data.} Our data includes the full population of more than 3 million instant messages sent/received and more than 1 million of trading decisions of day traders at a typical small-to-medium sized US trading company from 2007-2009. All trading related data was automatically captured by the company's trading system, which is specially designed for accuracy in recording, and used by most other companies in the industry. The study conforms to Institutional Review Board (IRB) criteria. There was no subject interaction, all data was $100\%$ archival, and the company and the subjects were anonymized. Legally, all data used in the study is owned by the company. All traders at the company know the company owns the data and that their communications and trading behavior is recorded by law. We received written permission from the company to use these data for research purposes and publishing contingent on identifying characteristics of the company and its traders remaining confidential and anonymous. For further details see Refs. \cite{Saavedra,Saavedra2}. We use only those tickers that we were able to disambiguate across traders' communications. Tickers' mentions and traded volume are normalized (to preserve the company's confidentiality) and provided in an additional file.
\\
\textbf{Control proximity matrices.} Partial Mantel correlations between network proximity and market similarity are controlled by proximity in market sector $\boldsymbol{F}$, average stock price in the year $\boldsymbol{T}$, board size $\boldsymbol{B}$, fraction of directors with financial expertise $\boldsymbol{E}$, and geographic distance $\boldsymbol{G}$. The proximity matrix $\boldsymbol{F}$ corresponds to $F_{i,j}=1$ if traded corporation $i$ and $j$ belong to the same sector, $0$ otherwise. In the rest of the control proximity matrices, the elements ($i,j$) of these matrices correspond to the inverse of the normalized absolute differences between the value measured in corporation $i$ and $j$, such that higher values always represent corporations with higher proximity. For proximity matrix $\boldsymbol{T}$, $T_{ij}=(\text{max}\{k1\}-k1_{ij})/\text{(max}\{k1\}-\text{min}\{k1\})$ where $k1_{ij}=|\langle \text{log}(p_i) \rangle - \langle \text{log}(p_j) \rangle|$ and $\langle \text{log}(p_i) \rangle$ is the average log price of stock $i$ over the year. For proximity matrix $\boldsymbol{B}$, $B_{ij}=(\text{max}\{k2\}-k2_{ij})/(\text{max}\{k2\}-\text{min}\{k2\})$ where $k2_{ij}=|b_i-b_j|$ and $b_i$ is the total number of board members of traded corporation $i$ over the year. For proximity matrix $\boldsymbol{E}$, $E_{ij}=(\text{max}\{k3\}-k3_{ij})/(\text{max}\{k3\}-\text{min}\{k3\})$ where $k3_{ij}=|e_i-e_j|$ and $e_i$ is the fraction of board members with financial expertise of traded corporation $i$ over the year. For proximity matrix $\boldsymbol{G}$, $G_{ij}=(\text{max}\{k4\}-k4_{ij})/(\text{max}\{k4\}-\text{min}\{k4\})$ where $k4_{ij}$ is the geographic distance between corporation $i$ and $j$ as function of their geographic coordinates latitude and longitude extracted from their zip codes. Zip codes were collected from {\it Yahoo Finance} and from {\it HighBeam.Com}. These data are provided in an additional file. We find a majority of negligible associations of market similarity with the control proximity matrices of firm size, board size, fraction of financial experts in the board, and geographic distance (see Additional Figs. S1-S4).
\\
\textbf{Robustness of correlations between network proximity and market similarity.} We find that each of the positive correlations shown in Figure 2 is also higher than the expected correlation between network proximity and market similarity when we randomly generate new matrices of network proximity (see Additional Fig. S5). These new matrices are generated by bootstrapping elements with replacement from the original matrices of network proximity. Additionally, the observed correlations between network proximity and market similarity shown in Fig. 2 cannot be reproduced if we disconnect boards (i.e., $d_{ij}=\infty$) when a certain degree of separation has been exceed (see Additional Fig. S6).  
\\
\textbf{Benchmark return.} To calculate the benchmark return, we use the average daily log returns $z_b= \langle z_i \rangle$ of all the observed corporations. For individual market sectors, we did not find statistical differences if we use $\langle z_i \rangle$ or the the average daily stock returns only of corporations from individual market sectors. The distribution of betas and yearly stock returns are shown in Additional Fig. S7.
\\
\textbf{Multivariate linear regression model} The model is defined by $\text{performance}_i  \sim \langle D_i \rangle + \langle \text{log}(p_i) \rangle + b_i + f_i + \epsilon_i$. The response variable $performance$ can take either the beta of the stock $\beta_i$ or the yearly stock return $r_i$. We control for the average price of the stocks in the year ($\langle \text{log}(p_i) \rangle$), the size of the boards ($b_i$), the fraction of directors with financial expertise in each board ($f_i$), and the Gaussian noise ($\epsilon_i$). We scale all predictor variables to be able to compare their effect. The model has a goodness-of-fit of R$^2 \sim 0.3$ for all years. Interestingly, the control variables board size and fraction of directors with financial expertise show a majority of negligible effects, while firm size yields positive and negative effects for short-term and long-term stock performance (see Additional Figs. S8-S10).
\\
\textbf{Mantel correlation} This correlation is the extension of the standard Pearson or Spearman rank correlation to dyadic data \cite{Mantel}. Mantel correlation works as follows, let us assume that we have two sets of $n$ objects represented by their similarity matrices $X_{ij}$ and $Y_{ij}$, i.e., dyadic data. For example in our case, network proximity and market similarity. The Pearson Mantel correlation coefficient is computed as
\begin{equation*}
r =  \frac{\sum_{i>j} (X_{ij} - \bar{X})(Y_{ij} - \bar{Y})}{\sum_{i>j} (X_{ij} - \bar{X})^2 \sum_{i>j} (Y_{ij} - \bar{Y})^2}
\end{equation*}
where $\bar{X} = \frac{2}{n(n-1)} \sum_{i>j} X_{ij}$ is the average $X{ij}$ value (similarly for $\bar{Y}$). Note that the sum is only made on the strictly upper triangular part of the matrices. This is so because similarity matrices are symmetric and have their diagonal elements equal to one. For the Spearman rank Mantel correlation, we first substitute the elements of the matrices $X_{ij}$ and $Y_{ij}$ by their respective rank. In a similar way we can compute partial Mantel correlations \cite{Legendre}. To compute the $95\%$ confidence interval of the correlation coefficient, we use a bootstrap procedure \cite{Bradley}. We re-sample simultaneously the rows and the columns of the similarity matrices, and then for each re-sampling we compute again the Mantel correlation coefficient. The re-sampling procedure results in the empirical distribution of the Mantel correlation coefficient, from which we extract the $95\%$ confidence interval.

\end{document}